\documentclass[12pt]{article}
\usepackage{graphicx}

\newcommand{\interlinia}{}

\title{Theory of the dielectric susceptibility
of liquid crystals with polar nonuniaxial molecules}

\author{A. Kapanowski \and T. Wietecha\\
{\em Institute of Physics, Jagellonian University,}\\
{\em ulica Reymonta 4, 30-059 Cracow, Poland}  }

\begin{document}
\maketitle

\interlinia

\begin{abstract}
\interlinia
Statistical theory of the dielectric susceptibility
of polar liquid crystals is proposed.
The molecules are not uniaxial but similar to cones.
It is assumed that the permanent dipole moment of a molecule
is parallel to the axis of the rotational symmetry.
The ordering of the phase is described by means of
the mean field theory based on the Maier-Saupe approach.
The theory was used to calculate the temperature
dependence of the order parameters and the susceptibilities.
Predictions of the model for different sets of parameters
are investigated.
\newline\newline
PACS number(s): 61.30.Cz, 77.84.Nh
\end{abstract}

\section{Introduction}
\label{sec1}

Liquid crystal phases consist of anisotropic molecules, 
either elongated or disk-like \cite{[1993_de_Gennes]}. 
In many theoretical 
\cite{[2000_Garcia],[1997_Zakhlevnykh]}
and computer simulation studies
\cite{[1992_TEAF],[1993_Zhang]}
it is assumed that the molecules
have the $D_{\infty h}$ (or $D_{2h}$) symmetry
and that the interactions do not change if a molecule
will point to the opposite direction.
However, real molecules are always less symmetrical.
Somoza and Tarazona
\cite{[1992_Somoza_Tarazona]}
showed that the molecular asymmetry can influence the packing
fraction, the relative density jump and the order parameter
at the isotropic-nematic transition for hard body systems.
Recently Barmes {\em et al}
\cite{[2003_Barmes]}
reported results from Monte Carlo simulations investigating
mesophase formation in systems of hard pear-shaped particles.
For different elongations isotropic, nematic,
interdigitated smectic $A_2$ phases were obtained and sometimes
glassy behaviour was seen.
A pear-shaped molecules was also used by Davidson and Mottram
\cite{[2002_Davidson_Mottram]}
in a continuum theory model of switching in a bistable menatic
liquid crystal device.

In the molecular-statistical theory of Maier and Saupe
the potential energy of interactions is
\begin{equation}
V(\vec{n}_1,\vec{n}_2) = v_0 + v_2 P_2(\vec{n}_1 \cdot \vec{n}_2),
\end{equation}
where $P_j$ are the Legendre polynomials, 
the vectors $\vec{n}_1,\vec{n}_2$ determine the orientation of molecules.
There is a nematic phase for $v_2 < 0$ ($v_0$ is not important).
We would like to investigate the interactions that depend on the
sense of the vectors $\vec{n}_1,\vec{n}_2$
(it corresponds the $C_{\infty v}$ symmetry of molecules)
\begin{equation}
V(\vec{n}_1,\vec{n}_2) = v_0 
+ v_1 P_1(\vec{n}_1 \cdot \vec{n}_2)
+ v_2 P_2(\vec{n}_1 \cdot \vec{n}_2).
\end{equation}
The parameters $v_i$ can be choosed by means of the excluded
volume method described by Straley
\cite{[1974_Straley]}. As a result we get always $v_1 \ge 0$.
The energy of interactions will be lower for $\vec{n}_1=-\vec{n}_2$
and this leads to the conclusion that only isotropic
and nematic phases can be obtained.

We can obtain negative (or positive) $v_1$ and new ordering of the phase
if we consider interactions of two electric dipols.
Under the assumption that $\vec{p}=p\vec{n}$, 
the interactions have the form
\begin{equation}
U_{12} (\vec{u},\vec{n}_1,\vec{n}_2) =
\frac{p^2 [(\vec{n}_1 \cdot \vec{n}_2)
-3(\vec{n}_1 \cdot \vec{\Delta})
(\vec{n}_2 \cdot \vec{\Delta})]}{4\pi \epsilon_0 u^3},
\end{equation}
where $\vec{u} = u \vec{\Delta}$ is the distance between dipols.
We would like to get average interactions that do not depend
on the vector $\vec{\Delta}$
\begin{equation}
U_{12}^{av} (u,\vec{n}_1,\vec{n}_2) = \frac{1}{4\pi}
\int {d\vec{\Delta}}
U_{12} (\vec{u},\vec{n}_1,\vec{n}_2)
g(\vec{\Delta},\vec{n}_1,\vec{n}_2),
\end{equation}
where the function $g(\vec{\Delta},\vec{n}_1,\vec{n}_2)$
describes the neighbourhood of the molecule 1 (how often
we can find the molecule 2 in the direction $\vec{\Delta}$
at the constant $u$).
The normalization is
\begin{equation}
\frac{1}{4\pi} \int {d\vec{\Delta}} 
g(\vec{\Delta},\vec{n}_1,\vec{n}_2) = 1.
\end{equation}
For isotropic distribution (spherical molecules)
$g(\vec{\Delta},\vec{n}_1,\vec{n}_2)=1$
and we get 
\begin{equation}
U_{12}^{av} (\vec{n}_1,\vec{n}_2) = 0.
\end{equation}
But for anisotropic molecules it is more reasonable
to try different form of the function
\begin{equation}
g(\vec{\Delta},\vec{n}_1,\vec{n}_2)
=1 - a P_2(\vec{n}_1 \cdot \vec{\Delta})
- a P_2(\vec{n}_2 \cdot \vec{\Delta}),
\end{equation}
where $a$ is related to the anisotropy of molecules:
rodlike (disclike) molecules have $a > 0$ ($a < 0$).
\begin{equation}
U_{12}^{av} (\vec{n}_1,\vec{n}_2) = 
\frac{p^2 (4/5)a (\vec{n}_1 \cdot \vec{n}_2)}{4\pi \epsilon_0 u^3}.
\end{equation}

Thus, for disclike molecules ($a<0$) one can use $v_1<0$
and this will produce ferroelectric ordering for sufficiently
low temperatures.
Generally, we can say that $v_1$ describes the result of sum of the excluded
volume effects and dipolar interactions.
Our aim is to investigate the influence of the $v_1$ term on the ordering
and the dielectric susceptibility of the phase.

As far as the susceptibility is concerned, we would like to recall
some results in order to provide the context for our considerations.
The potential energy of the dipole $\vec{p}$ 
in the electric field $\vec{E}$ is
$U = - \vec{p} \cdot \vec{E}$.
For a system of $N$ noninteracting dipole moments in the volume $V$ 
we get the polarization $P$ from the Boltzmann distribution 
\begin{equation}
P = \frac{Np}{V} L\left( \beta pE \right),
\end{equation}
where $\beta=1/(k_B T)$ and
$L(x)=\coth x-1/x$ is the Langevin function.
For small values of the argument, $L(x)\approx x/3$, and thus
\begin{equation}
P = \frac{N p^2 \beta E}{3 V}.
\end{equation}
The susceptibility is defined as
\begin{equation}
\epsilon_0 \chi = \frac{\partial P}{\partial E} 
= \frac{N p^2}{3 V k_B T}.
\end{equation}
We get a $1/T$ dependence called the Curie law. It indicates
that the dipoles are noninteracting.

For the anisotropic polar molecules and anisotropic phases
the susceptibility is a tensor
\begin{equation}
\label{chiab}
\epsilon_0 \chi_{\alpha\beta} = 
\frac{\partial P_{\alpha}}{\partial E_{\beta}}.
\end{equation}
The polarization can be written as
\begin{equation}
P_{\alpha} = \frac{Np}{V} \langle \bar{n}_{\alpha} \rangle,
\end{equation}
where $p\langle \bar{n}_{\alpha} \rangle$ is the average value of the dipole
component in the $\alpha$-direction in the presence of the electric field.
Hence two different averages are involved. The bar refers to the potential
energy of the dipole moment in the electric field and the brakets to the 
nematic potential.
We calculate it using the linear approximation
\begin{equation}
\langle \bar{n}_{\alpha} \rangle = 
\langle n_{\alpha} (1+\beta p n_{\beta} E_{\beta}) \rangle =
\beta p \langle n_{\alpha} n_{\beta} \rangle E_{\beta}.
\end{equation}
The susceptibility is
\begin{equation}
\epsilon_0 \chi_{\alpha\beta} = 
\frac{N p^2 \beta}{V} \langle n_{\alpha} n_{\beta} \rangle.
\end{equation}
The components of the susceptibility in the nematic phase
oriented along the z axis are
\begin{eqnarray}
\epsilon_0 \chi_{zz} &=& \frac{N p^2 (1+2S)}{3V k_B T}, \\ 
\epsilon_0 \chi_{xx}&=& \frac{N p^2 (1-S)}{3V k_B T},
\end{eqnarray}
where $S=\langle P_2(n_z) \rangle$ is the order parameter.
Note that the average susceptibility is
\begin{equation}
\bar{\chi} 
\equiv \frac{1}{3} (\chi_{zz}+2\chi_{xx})
=\frac{N p^2\beta}{3V\epsilon_0} = \chi^{iso}
\end{equation}
as for the isotropic phase. 
The equality $\bar{\chi}=\chi^{iso}$ is correct for nonpolar molecules
but for polar ones we know from the experiments that
$\bar{\chi} < \chi^{iso}$.
More advanced calculations were given by Maier and Meier
\cite{[1961_Maier_Meier]}
who extended the Onsager theory of the susceptibility
to nematic liquid crystals. If we neglect the induced polarization
we can write their results as
\begin{eqnarray}
\epsilon_0 \chi_{zz} &=& 
\left(\frac{3 \tilde{\chi} + 3}{2 \tilde{\chi} +3} \right)
\frac{N p^2 (1+2S)}{3V k_B T}, \\ 
\epsilon_0 \chi_{xx}&=& 
\left(\frac{3 \tilde{\chi} + 3}{2 \tilde{\chi} +3} \right)
\frac{N p^2 (1-S)}{3V k_B T},
\end{eqnarray}
where $\tilde{\chi}$ is the average susceptibility of the phase.

The organization of this paper is as follows:
In Sec. \ref{sec2} the mean field theory of the nematic ordering
is provided.
In Sec. \ref{sec3} the formulae for the susceptibility are derived.
Section \ref{sec4} is devoted to some applications
of the presented theory.
Section \ref{sec5} contains a summary.
Appendixes provide the definitions and main properties
of the basic functions (App.~A)
and the excluded volumes for molecules of different shape
(App.~B).

\section{The mean field theory}
\label{sec2}

The potential energy of molecular interactions of the molecules
is given by
\begin{equation}
V(R_1,R_2) = 
\sum_{j=0,1,2} v_j E_{00}^{(j)}(R_2^{-1} R_1),
\end{equation}
where $R_i$ are the sets of the three Euler angles describing the orientation
of the molecule $i$, $E_{\mu\nu}^{(j)}$ are the basic functions 
defined in App.~A.
The molecular orientation in the phase is described by the distribution 
function
\begin{equation}
\int \! {dR} f(R) = 1. 
\end{equation}
The mean of some function $A=A(R)$ we calculate as
\begin{equation}
\langle A \rangle  \equiv  \int \! {dR} f(R) A(R).
\end{equation}
The state of the system is described by order parameters
$\langle E_{\mu0}^{(j)} \rangle$ but the most important is
$S=\langle E_{00}^{(2)} \rangle$.
The internal energy of the system is
\begin{equation}
U = \frac{N}{2} \int \! {dR_1} {dR_2} f(R_1) f(R_2) V(R_1,R_2),
\end{equation}
whereas the entropy of the system is written as
\begin{equation}
S = -k_B N \int \! {dR} f(R) \ln [f(R) C_N].
\end{equation}
The energy of permanent dipole moments $\vec{p}=p\vec{n}$
in the internal electric field $\vec{E}^{int}$ is
\begin{equation}
U^E =  N \langle \vec{p}  \cdot \vec{E}^{int} \rangle.
\end{equation}
Generally, the internal electric field differs from
the external field $\vec{E}^{ext}$
\cite{[1978_Jeu]}.
Here we assume that both field are equal
$\vec{E}^{int} = \vec{E}^{ext} = \vec{E}$.

The total free energy of the system is the sum
\begin{equation} F^{tot}=U+U^E-TS. \end{equation}
In the mean-field approximation we get a potential energy
\begin{equation}
W(R) = \sum_{j} \sum_{\mu}
w_{\mu0}^{(j)} E_{\mu0}^{(j)}(R),
\end{equation}
and from the Boltzmann distribution we get
\begin{equation}
f(R) = \exp [-\beta W(R)]/Z,
\end{equation}
where $Z$ is a normalization constant.
The consistency condition
\begin{equation}
W(R_1) =  \int \! {dR_2} f(R_2) V(R_1,R_2)
-\vec{p}(R_1) \cdot \vec{E}
\end{equation}
leads to equations
\begin{equation}
w_{\mu0}^{(j)} = \langle E_{\mu0}^{(j)} \rangle
v_j -p \delta_{1j} 
(E_x \delta_{1\mu}-E_y \delta_{-1\mu}+E_z \delta_{0\mu}).
\end{equation}
It is useful to introduce the dimensionless parameters
$S_{\mu0}^{(j)}= -\beta w_{\mu0}^{(j)}$ for $j>0$
\begin{equation}
\ln f(R) = \sum_j \sum_{\mu} S_{\mu0}^{(j)} E_{\mu0}^{(j)}(R).
\end{equation}
$S_{00}^{(0)}$ is responsible for the normalization and it depends
on other $S_{\mu0}^{(j)}$ with $j>0$
\begin{equation} 
S_{00}^{(0)}= -\ln \left[\int {dR}
\exp\left(\sum_{j>0} \sum_{\mu} 
S_{\mu0}^{(j)} E_{\mu0}^{(j)}(R)\right) \right] ,
\end{equation}
\begin{equation}
\frac{\partial S_{00}^{(0)}}{\partial S_{\mu0}^{(j)}}
= -\langle E_{\mu0}^{(j)} \rangle,
\end{equation}
\begin{equation}
W_{\mu\nu}^{jk}\equiv
\frac{\partial \langle E_{\mu0}^{(j)}\rangle}{\partial S_{\nu0}^{(k)}}
=\langle E_{\mu0}^{(j)}E_{\nu0}^{(k)} \rangle 
- \langle E_{\mu0}^{(j)} \rangle \langle E_{\nu0}^{(k)} \rangle .
\end{equation}
The equations have the form
\begin{equation}
S_{\mu0}^{(j)} + \beta v_j \langle E_{\mu0}^{(j)} \rangle
= \beta p \delta_{1j} 
(E_x \delta_{1\mu}-E_y \delta_{-1\mu}+E_z \delta_{0\mu}).
\end{equation}
The solution is orientationally stable only if the matrix
\begin{equation}
\left[
W_{\mu\nu}^{jk} + \sum_{l>0} \sum_{\rho} \beta v_l 
W_{\mu\rho}^{jl} W_{\nu\rho}^{kl}
\right]
\end{equation}
is positive definite. The isotropic phase is orientationally stable
if $\beta v_1 > -3$ 
and $\beta v_2 > -5$.

\section{The dielectric susceptibility}
\label{sec3}

The dielectric susceptibility tensor is defined by the eq. 
(\ref{chiab}) and the orientational polarization
we calculate as
\begin{equation}
P_{\alpha} = \langle n_{\alpha} \rangle N p/V.
\end{equation}
Note that the polarization depends on the electric field via the
distribution function.
The components of the susceptibility are
\begin{eqnarray}
\epsilon_0 \chi_{zz} &=& 
\frac{N p}{V} \sum_{j>0}\sum_{\mu} W_{0\mu}^{1j}
\frac{\partial S_{\mu 0}^{(j)}}{\partial E_z}, \\
\epsilon_0 \chi_{xx} &=& 
\frac{N p}{V} \sum_{j>0}\sum_{\mu} W_{1\mu}^{1j}
\frac{\partial S_{\mu 0}^{(j)}}{\partial E_x},
\end{eqnarray}
where the derivatives are calculated from the equations
\begin{eqnarray}
\sum_{k>0}\sum_{\nu}\left[ \delta_{jk}\delta_{\mu\nu}
+\beta v_j W_{\mu\nu}^{jk} \right]
\frac{\partial S_{\nu 0}^{(k)}}{\partial E_z} 
&=& \beta p \delta_{1j} \delta_{0\mu}, 
\\
\sum_{k>0}\sum_{\nu}\left[ \delta_{jk}\delta_{\mu\nu}
+\beta v_j W_{\mu\nu}^{jk} \right]
\frac{\partial S_{\nu 0}^{(k)}}{\partial E_x} 
&=& \beta p \delta_{1j} \delta_{1\mu}.
\end{eqnarray}
Now we are in the position to disscuss the results for
different possible phases.

\subsection{The isotropic phase}

For the zero field all order parameters are equal to zero.
For the nonzero field the phase has the symmetry 
$C_{\infty v}$ (the symmetry of the electric field).
Thus we have to use the parameters $S_{00}^{(j)}$ only.
\begin{equation}
\epsilon_0^{iso} 
= \epsilon_0 \chi_{zz} 
= \epsilon_0 \chi_{xx} 
= \frac{N p^2}{V(3 k_B T + v_1)}.
\end{equation}
For $v_1<0$ we get the Curie-Weiss law 
describing the divergence of $\chi$
when we approach the Curie temperature from above.
For $v_1>0$ the susceptibility is finite.

\subsection{The uniaxial nematic phase}

For the zero field the phase has the symmetry $D_{\infty h}$ and
the parameters $S_{00}^{(j)}$ with $j$ even are present.
For the parallel field the phase has the symmetry $C_{\infty v}$ and
the parameters $S_{00}^{(j)}$ with all $j$ are present.
For the perpendicular field the phase has the symmetry $C_{2 v}$ and
we have the parameters $S_{00}^{(j)}$ with $j$ even and
$S_{\mu0}^{(j)}$ with $j,\mu$ odd.
The expressions $W_{\mu\nu}^{jk}$ with $j+k$ and $\mu+\nu$ even are nonzero.
\begin{eqnarray}
\epsilon_0 \chi_{zz} &=& 
\frac{N p^2 \beta W_{00}^{11}}{V (1+\beta v_1 W_{00}^{11})} ,
\\
\epsilon_0 \chi_{xx} &=& 
\frac{N p^2 \beta W_{11}^{11}}{V (1+\beta v_1 W_{11}^{11})}.
\end{eqnarray}
Note that for $v_1>0$ we get $\bar{\chi} < \chi^{iso}$.

\subsection{The ferroelectric phase}

For the zero field and for the parallel field the phase has 
the symmetry $C_{\infty v}$ and the parameters $S_{00}^{(j)}$ 
with all $j$ are present.
The expressions $W_{\mu\nu}^{jk}$ with $\mu+\nu$ even are nonzero.
\begin{equation}
\epsilon_0 \chi_{zz} =
\left( \frac{N p^2 \beta}{V} \right)
\frac{ W_{00}^{11} (1+\beta v_2 W_{00}^{22})
- \beta v_2 (W_{00}^{12})^2}
{(1+\beta v_1 W_{00}^{11})(1+\beta v_2 W_{00}^{22})
-\beta v_1 \beta v_2 (W_{00}^{12})^2} .
\end{equation}
For convenience, we express the important elements
$W_{\mu\nu}^{jk}$ by means of the order parameters
$\langle P_j \rangle$ in the case of the ferroelectric phase
\begin{eqnarray}
W_{00}^{11} & = & [1+2 \langle P_2 \rangle]/3 - \langle P_1 \rangle^2 ,\\
W_{00}^{12} & = & [2\langle P_1 \rangle + 3\langle P_3 \rangle]/5
-\langle P_1 \rangle\langle P_2 \rangle,\\
W_{00}^{22} & = & [7+10\langle P_2 \rangle +18\langle P_4 \rangle]/35
-\langle P_2 \rangle^2,\\
W_{11}^{11} & = & [1-\langle P_2 \rangle]/3  ,\\
W_{11}^{12} & = & [\langle P_1 \rangle -\langle P_3 \rangle]\sqrt{3}/5  ,\\
W_{11}^{22} & = & [7+5\langle P_2 \rangle -12\langle P_4 \rangle]/35  ,\\
W_{20}^{22} & = &  0 ,\\
W_{22}^{22} & = &  [7-10\langle P_2 \rangle +3\langle P_4 \rangle]/35 ,\\
W_{02}^{12} & = &  0.
\end{eqnarray}

\section{Exemplary calculations}
\label{sec4}

In this section we carry out calculations for different
physical systems of polar molecules described 
by the considered model.
The phase diagram of the model is shown in Fig. 1.
Three phases are present: isotropic, uniaxial nematic
and ferroelectric.
Dashed half-lines in the picture denote different
physical systems with the fixed parameters $v_1$ and $v_2$ 
(in fact, the fixed ratio $v_2$ to $v_1$ is important).
The parameters of the model were calculated by means of the
excluded volume method.
The formulae for molecules with different shapes
are given in App. B.

Let us start the description of the results from the
system of uniaxial molecules similar to ellipsoids
or cylinders [half-line (d) in Fig. 1, $v_1=0$].
In all pictures, $T$ denotes the dimensionless temperature.
$T=1$ corresponds to the transition from the isotropic
to the nematic or ferroelectric phase.
The susceptibilities are scaled in order to obtain
$\chi=1$ for $T=1$.
The temperature dependence of the order parameter
$\langle P_2 \rangle$ and the susceptibilities
are presented in Figs. 2 and 3, respectively.
There is the first order transition from the isotropic
to the nematic phase.
On decreasing the temperature the susceptibility
splits into $\chi_{xx} < \chi_{zz}$. $\chi_{zz}$
runs to the infinity whereas $\chi_{xx}$ remains finite.
This is typical for the uniaxial molecules with the dipol
moment parallel to the symmetry axis.

Let us consider less symmetrical molecules similar to cones
with $H/D=4$, where $H$ and $D$ denote the height and
the diameter of the base, respectively
[half-line (e) in Fig. 1, $v_2= -6 v_1$].
Note that only the isotropic and nematic phases are present.
This reflects the fact that liquid crystals are in general
nonpolar. On the other hand, one should expect
the flexoelectric effect which is polar
\cite{[1969_Meyer]}.
The temperature dependence of the order parameter
$\langle P_2 \rangle$  is similar to shown in Fig. 2.
The temperature dependence of the susceptibilities is presented
in Fig. 4. We have known splitting of the susceptibility
but here both $\chi_{xx}$ and $\chi_{zz}$ are finite.
In Fig. 5 the results for $v_2=-v_1$ are given
[half-line (f) in Fig. 1] where $\chi_{xx}$ and $\chi_{zz}$
are almost the same order. Note that this value of the $v_2/v_1$ 
cannot be obtained from the excluded volume method but we use it
in order to understand the behaviour of the model.
We conclude that 
if cones become longer then their properties are more
similar to ellipsoids and cylinders.
For shorter cones it is more difficult for the electric field
to rotate the molecule and that is why the susceptibility
$\chi_{zz}$ is lower (but still greater then $\chi_{xx}$).

A ferroelectric phase is present in the considered model
when the parameter $v_1$ is negative and there is
a resultant tendency to arrange the dipols in a parallel
fashion.
The temperature dependence of the order parameters
and the susceptibilities
in the case of $v_2=5v_1$ [half-line (c) in Fig. 1]
are presented in Figs. 6 and 7, respectively.
There is the first order transition from the isotropic
to the nematic phase at $T=1$. Next, there is the second order
transition to the ferroelectric phase at $T=0.763$.

The temperature dependence of the order parameters
and the susceptibilities
in the case of $v_2=v_1$ [half-line (b) in Fig. 1]
are presented in Figs. 8 and 9, respectively.
There is the first order transition from the isotropic
to the ferroelectric phase.
$\chi_{zz}$ is always finite and it has the strong maximum at the transition.

For the case of $v_2=0$ we get the simple equation for $S_{00}^{(1)}$ only
[half-line (a) in Fig. 1]
\begin{equation}
S_{00}^{(1)}+\beta v_1 L(S_{00}^{(1)})=\beta p E_z.
\end{equation}
In the zero field there is the second order transition from the isotropic 
to the ferroelectric phase at $T_C= -v_1/(3k_B)$.
The order parameters shown in Fig. 10 can be expressed as
\begin{eqnarray}
\langle P_1 \rangle & = & L(S_{00}^{(1)}), \\
\langle P_2 \rangle & = & 1-T/T_C.
\end{eqnarray}
As far as the susceptibility is concerned, we recover a general 
mean field result: the susceptibility in the ordered phase 
is half of that in the disordered phase 
(at the neighbourhood of the transition point)
\cite{[handbook]}.
The reason is that the effect of an external field on the polarization
must be smaller in the ordered phase, due to the existing internal field
in that phase, which has a stabilizing effect on the polarization.
In Fig. 11 we plot the inverse of the susceptibility as a function
of the temperature. The dashed line shows the results
from the Landau description for the ferroelectric phase.
We have an expected agreement in the neighbourhood of the transition point.

\section{Summary}
\label{sec5}

In this paper we considered the mean field theory 
with interactions which are not uniaxial.
We showed that they correspond to molecules similar to cones
or molecules with permanent dipol moments parallel to the axis
of rotational symmetry. Three different phases were obtained:
isotropic (I), nematic (N) and ferroelectric (F).
We found the first (I-N, I-F) and
the second order transitions (N-F, I-F)
for some sets o parameters.

We also derived the formulae for the dielectric susceptibility
in the case of polar molecules.
We neglected the induced polarization and 
the difference between the external electric field and the internal
electric field acting on a molecule.
The susceptibilities depend on the density, square of the dipolar moment, 
temperature and the order parameters.
The anisotropy $\Delta \chi=\chi_{zz}-\chi_{xx}$ is positive
as a result of the assumption on the dipol moment orientation.

Our calculations show that for positive or small negative
values of $v_1$ the ordering is the same as in the case
of $v_1=0$ (uniaxial molecules). 
Below the isotropic phase there is the nematic phase
with the $D_{\infty h}$ symmetry.
On the other hand, the nonzero $v_1$ strongly changes 
the component of the susceptibility which is parallel
to the symmetry axis of the molecule.
We also find that for $v_1 > 0$ the average susceptibility
is lower than the isotropic susceptibility what is
in agreement with the experiments for polar molecules.

The presented model describes only liquid crystal phases 
with translational symmetry. 
For real liquid crystals below the nematic phase
(for lower temperatures)
one can usually find the smectic phase translationally ordered
in one direction (rod-like molecules)
or the columnar phase translationally ordered in two directions
(disc-like molecules)
\cite{[1990_Chandrasekhar_Ranganath]}.
Sometimes the nematic phase is even absent.
It is possible to extend our model for smectic phases
by extension of the McMillan model
\cite{[1971_McMillan]}
or to include the columnar phases by extension of the model
by Feldkamp {\em et al}
\cite{[1981_Feldkamp_Handschy_Clark]}.
Work is in progress on the theory were many limitations of the presented 
approach is removed, i.e. any orientation of the dipol moment is possible,
the induced polarization and difference between external and
internal electric fields are taken into account.

\section*{Appendix A}
\label{app1}

Below we list the properties of the functions $E_{\mu\nu}^{(j)}$.
The functions can be used to describe any physical quantity
which depends on the three Euler angles.

\begin{enumerate}
\item 
The definition is 
\begin{eqnarray}
& E_{\mu\nu}^{(j)}(R) = 
\left( 
{\frac{1}{\sqrt{2}}} 
\right)^{2+\delta_{0\mu}+\delta_{0\nu}}
{\frac{1}{2}}
[(1+i)+(1-i) \mbox{sign}(\mu) \mbox{sign}(\nu)]
& \nonumber\\
& \times [ D_{\mu\nu}^{(j)}(R)
+ \mbox{sign}(\mu) \mbox{sign}(\nu) (-1)^{\mu+\nu} 
D_{-\mu,-\nu}^{(j)}(R) 
& \nonumber\\
& \mbox{} + \mbox{sign}(\nu) (-1)^{\nu} D_{\mu,-\nu}^{(j)}(R)
+ \mbox{sign}(\mu) (-1)^{\mu} D_{-\mu,\nu}^{(j)}(R) ], &
\end{eqnarray}
where $R=(\phi,\theta,\psi)$ (the three Euler angles),
$j$ is a non-negative integer, 
$\mu$ and $\nu$ are integers.
Functions $D_{\mu\nu}^{(j)}$ are 
standard rotation matrix elements \cite{[1957_Edmonds]} and
\begin{equation}
\mbox{sign}(x)=
\left\{
\begin{array}{rl}
 1 & \mbox{for}\ x \ge 0 \\ 
 -1 & \mbox{for}\ x < 0. 
\end{array}
\right.
\end{equation}
Note that
\begin{equation} 
\mbox{sign}(-x) = - \mbox{sign}(x) + 2\delta_{0x}. 
\end{equation}
\item 
The functions $E_{\mu\nu}^{(j)}$ are real.
\item
The functions satisfy the orthogonality relations
\begin{equation}
\int \! {dR} E_{\mu\nu}^{(j)}(R) E_{\rho\sigma}^{(k)}(R)=
\delta_{jk} \delta_{\mu\rho} \delta_{\nu\sigma}
8\pi^{2}/(2j+1).
\end{equation}
\item
Let us assume that the three Euler angles $R=(\phi,\theta,\psi)$
determine the orientation of the three unit orthogonal vectors
$(\vec{l},\vec{m},\vec{n})$. The coordinates of the vectors can be
expressed by means of the functions $E_{\mu\nu}^{(1)}$
\begin{eqnarray}
& l_{x} = E_{11}^{(1)}(R) 
= \cos\theta\cos\phi\cos\psi-\sin\phi\sin\psi, & \nonumber\\
& l_{y} = -E_{-11}^{(1)}(R) 
= \cos\theta\sin\phi\cos\psi+\cos\phi\sin\psi, & \nonumber\\
& l_{z} = E_{01}^{(1)}(R) 
= -\sin\theta\cos\psi, &  \nonumber\\
& m_{x}  = E_{1-1}^{(1)}(R) 
= -\cos\theta\cos\phi\sin\psi-\sin\phi\cos\psi, &  \nonumber\\
& m_{y} = E_{-1-1}^{(1)}(R) 
= -\cos\theta\sin\phi\sin\psi+\cos\phi\cos\psi, &  \nonumber\\
& m_{z} = E_{0-1}^{(1)}(R) = \sin\theta\sin\psi, &  \nonumber\\
& n_{x} = E_{10}^{(1)}(R) = \sin\theta\cos\phi, &  \nonumber\\
& n_{y} = -E_{-10}^{(1)}(R) = \sin\theta\sin\phi, &  \nonumber\\
& n_{z} = E_{00}^{(1)}(R) = \cos\theta.&   
\end{eqnarray}
\item
The functions $E_{\mu0}^{(j)}$ are proper for the description
of uniaxial molecules. We can express them as products
$n_x^a n_y^b n_z^c$, 
where $b=0$ or $b=1$.
\begin{eqnarray}
E_{00}^{(   0   )} & = & 1, \nonumber\\
E_{10}^{(   1   )} & = & n_x, \nonumber\\
E_{00}^{(   1   )} & = & n_z, \nonumber\\
E_{-10}^{(   1   )} & = & -n_y, \nonumber\\
E_{20}^{(   2   )} & = & 1/2 (-1+{n_z}^{2}+2{n_x}^{2} )\sqrt {3}, \nonumber\\
E_{10}^{(   2   )} & = & n_x n_z\sqrt {3}, \nonumber\\
E_{00}^{(   2   )} & = & -1/2+3/2{n_z}^{2}, \nonumber\\
E_{-10}^{(   2   )} & = &-n_z n_y\sqrt {3}, \nonumber\\
E_{-20}^{(   2   )} & = & n_x n_y\sqrt {3}, \nonumber\\
E_{30}^{(   3   )} & = &
1/4n_x (-3+3{n_z}^{2}+4{n_x}^{2} )\sqrt {5}\sqrt {2}, \nonumber\\
E_{20}^{(   3   )} & = &
1/2n_z (-1+{n_z}^{2}+2{n_x}^{2} )\sqrt {3}\sqrt {5}, \nonumber\\
E_{10}^{(   3   )} & = &
1/4n_x (-1+5{n_z}^{2} )\sqrt {3}\sqrt {2}, \nonumber\\
E_{00}^{(   3   )} & = & -3/2n_z+5/2{n_z}^{3}, \nonumber\\
E_{-10}^{(   3   )} & = &
-1/4n_y (-1+5{n_z}^{2} )\sqrt {3}\sqrt {2}, \nonumber\\
E_{-20}^{(   3   )} & = &
n_x n_z n_y\sqrt {3}\sqrt {5}, \nonumber\\
E_{-30}^{(   3   )} & = &
-1/4 (-1+{n_z}^{2}+4{n_x}^{2} )n_y \sqrt {5}\sqrt {2}, \nonumber\\
E_{40}^{(   4   )} & = &
1/8 (1-2{n_z}^{2}-8{n_x}^{2}+{n_z}^{4}+8
{n_x}^{2}{n_z}^{2}+8{n_x}^{4} )\sqrt {35}, \nonumber\\
E_{30}^{(   4   )} & = &
1/4n_x n_z (-3+3{n_z}^{2}+4{n_x}^{2} )\sqrt {35}\sqrt {2}, \nonumber\\
E_{20}^{(   4   )} & = &
1/4 (1-8{n_z}^{2}-2{n_x}^{2}+7{n_z}^{4}+
14{n_x}^{2}{n_z}^{2} )\sqrt {5}, \nonumber\\
E_{10}^{(   4   )} & = &
1/4 n_x n_z (-3+7{n_z}^{2} )\sqrt {5}\sqrt {2}, \nonumber\\
E_{00}^{(   4   )} & = &
3/8-{\frac {15}{4}}{n_z}^{2}+{\frac {35}{8}}{n_z}^{4}, \nonumber\\
E_{-10}^{(   4   )} & = &
-1/4n_z n_y (-3+7{n_z}^{2} )\sqrt {5}\sqrt {2}, \nonumber\\
E_{-20}^{(   4   )} & = &
1/2n_x n_y (-1+7{n_z}^{2} )\sqrt {5}, \nonumber\\
E_{-30}^{(   4   )} & = &
-1/4n_z n_y (-1+{n_z}^{2}+4{n_x}^{2} )\sqrt {35}\sqrt {2}, \nonumber\\
E_{-40}^{(   4   )} & = &
1/2n_x n_y (-1+{n_z}^{2}+2{n_x}^{2} )\sqrt {35}. \nonumber
\end{eqnarray}
\end{enumerate}

\section*{Appendix B}
\label{app2}

Excluded volumes $V_{excl}$ for uniaxial molecules with
the $D_{\infty h}$ and $C_{\infty v}$ symmetry.
$V_{mol}$ denotes the volume of a molecule.
The excluded volumes were used to calculate the parameters
of the interactions $v_j$ ($j=0,1,2$) from the expressions
\begin{eqnarray}
V_{excl}(\vec{e}_z,\vec{e}_z) & \mbox{is proportional to} & v_0+v_1+v_2, \\ 
V_{excl}(\vec{e}_z,-\vec{e}_z) & \mbox{is proportional to} & v_0-v_1+v_2, \\ 
V_{excl}(\vec{e}_z,\vec{e}_x) & \mbox{is proportional to} & v_0-v_2/2.
\end{eqnarray}

\begin{enumerate}
\item
Ellipsoids with the axes $2a \times 2a \times 2c$
($D_{\infty h}$ symmetry).
\begin{eqnarray}
V_{mol} & = & (4/3)\pi a^2 c, \\
V_{excl}(\vec{e}_z,\vec{e}_z) & = & 8 V_{mol}, \\ 
V_{excl}(\vec{e}_z,\vec{e}_x) & \approx & (4/3)\pi (a+c)^2 (2a) .
\end{eqnarray}
\item
Cylinders with the length $H$ and the diameter $2R$
($D_{\infty h}$ symmetry).
\begin{eqnarray}
V_{mol} & = & \pi R^2 H, \\
V_{excl}(\vec{e}_z,\vec{e}_z) & = & 8 V_{mol}, \\ 
V_{excl}(\vec{e}_z,\vec{e}_x) & = & 4\pi R^3 + 2\pi R^2 H + 4 R H^2.
\end{eqnarray}
\item
Cones with the height $H$ and the diameter of the base $2R$
($C_{\infty v}$ symmetry).
\begin{eqnarray}
V_{mol} & = & (1/3)\pi R^2 H, \\
V_{excl}(\vec{e}_z,\vec{e}_z) & = & 14 V_{mol}, \\ 
V_{excl}(\vec{e}_z,-\vec{e}_z) & = & 8 V_{mol}, \\ 
V_{excl}(\vec{e}_z,\vec{e}_x) & \approx & 12 R^3 + 8 R^2 H + 4 R H^2.
\end{eqnarray}
\item
"Towers" that consist of two cylinders which 
have the common axis of rotational symmetry ($C_{\infty v}$ symmetry).
The cylinders are of the height $H/2$
and the diameters $R$ and $2R$. 
\begin{eqnarray}
V_{mol} & = & (5/8)\pi R^2 H, \\
V_{excl}(\vec{e}_z,\vec{e}_z) & = & 10 V_{mol}, \\ 
V_{excl}(\vec{e}_z,-\vec{e}_z) & = & 9 V_{mol}, \\ 
V_{excl}(\vec{e}_z,\vec{e}_x) & \approx & 15 R^3 + 12 R^2 H + 4 R H^2.
\end{eqnarray}
\end{enumerate}

\pagebreak

\begin{figure}
\begin{center}
\includegraphics{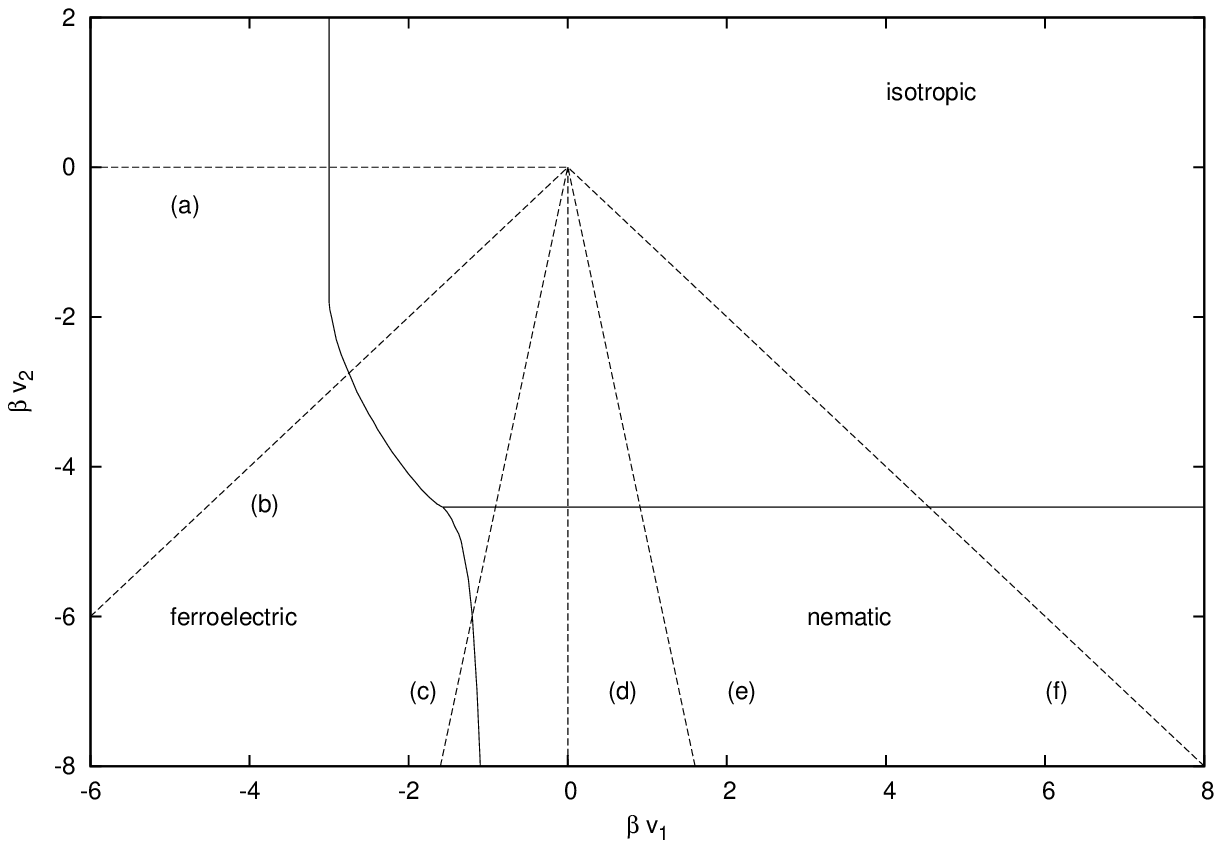}
\end{center}
\caption[Phase diagram of the model.]{
\label{fig1}
\interlinia
Phase diagram of the model considered in the paper.
Three phases are present: isotropic, uniaxial nematic
and ferroelectric.
Dashed half-lines denote different physical systems:
(a) $v_2=0$,
(b) $v_2=v_1$,
(c) $v_2=5 v_1$,
(d) $v_1=0$,
(e) $v_2=-6 v_1$ and
(f) $v_2=-v_1$.
For a given physical system on decreasing the temperature
we are moving from the center (0,0) to the edge of the figure.}
\end{figure}

\begin{figure}
\begin{center}
\includegraphics{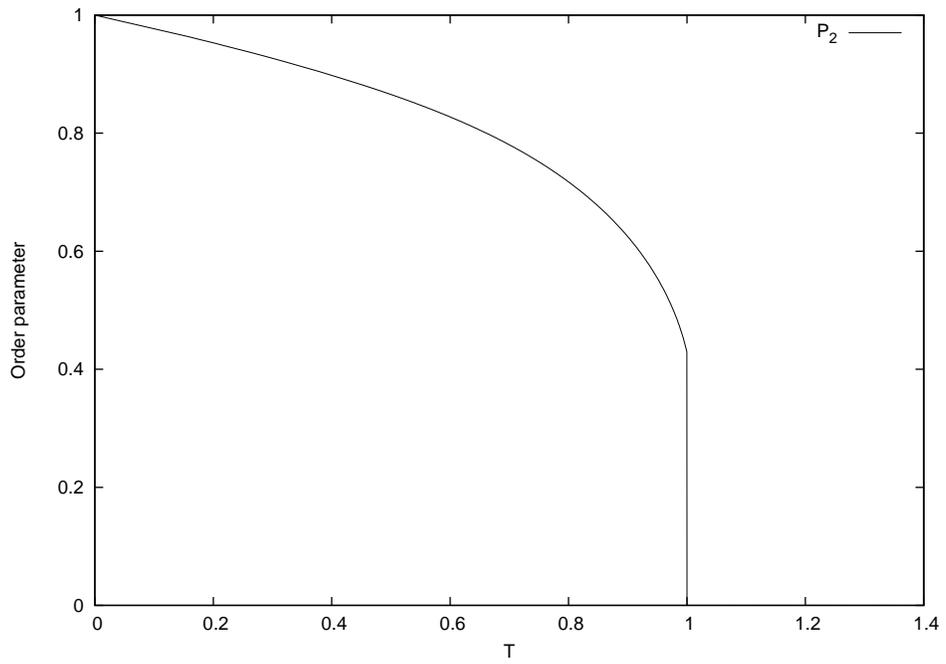}
\end{center}
\caption[Temperature dependence of the order parameter 
for $v_1=0$.]{
\label{fig2}
\interlinia
Temperature dependence of the order parameter
$\langle P_2 \rangle$ for $v_1=0$ [half-line (d) in Fig. 1].
There is first order transition from the isotropic to
the uniaxial nematic phase.
$T$ denotes the dimensionless temperature.}
\end{figure}

\begin{figure}
\begin{center}
\includegraphics{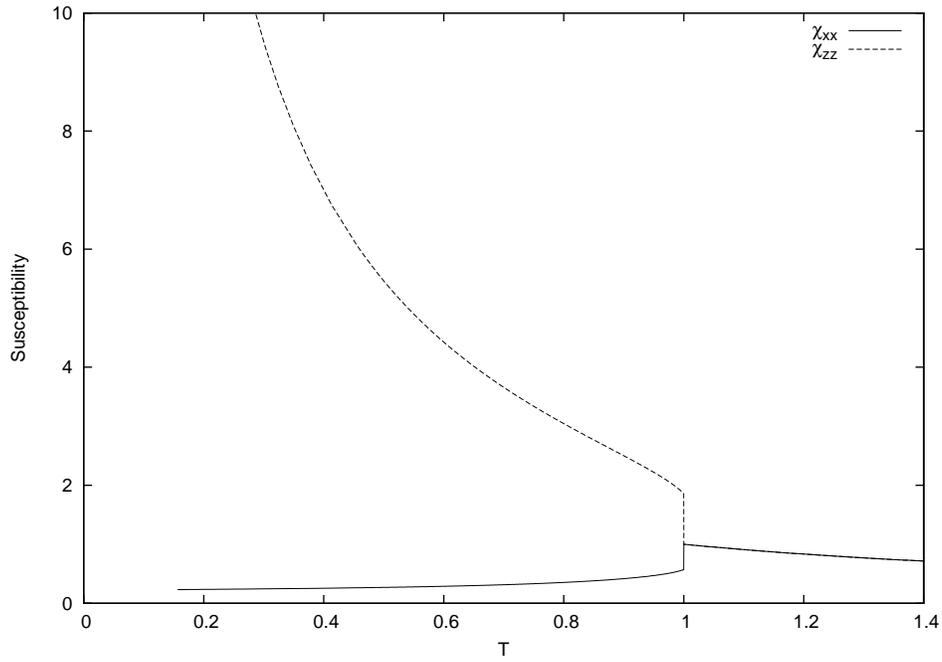}
\end{center}
\caption[Temperature dependence of the susceptibilities
for $v_1 = 0$.]{
\label{fig3}
\interlinia
Temperature dependence of the susceptibilities
for $v_1=0$ [half-line (d) in Fig. 1].
On decreasing the temperature the susceptibility
splits into $\chi_{xx} < \chi_{zz}$.
For $T\rightarrow 0$ $\chi_{zz}$ becomes infinite.
$T$ denotes the dimensionless temperature.}
\end{figure}

\begin{figure}
\begin{center}
\includegraphics{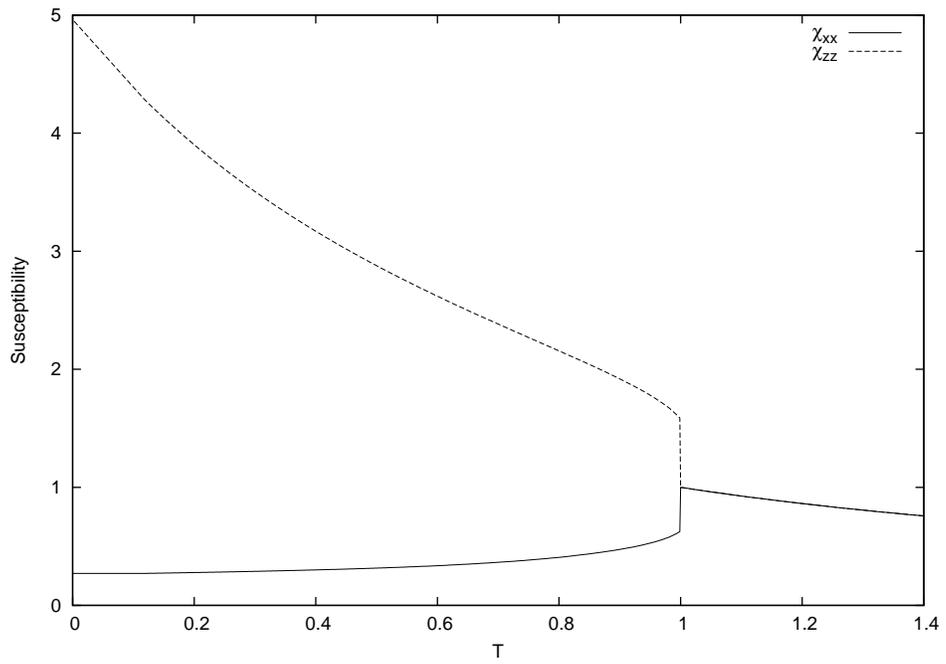}
\end{center}
\caption[Temperature dependence of the susceptibilities
for $v_2 = -6 v_1$.]{
\label{fig4}
\interlinia
Temperature dependence of the susceptibilities
for $v_2= -6 v_1$ [half-line (e) in Fig. 1].
Both susceptibilities are finite.
$T$ denotes the dimensionless temperature.}
\end{figure}

\begin{figure}
\begin{center}
\includegraphics{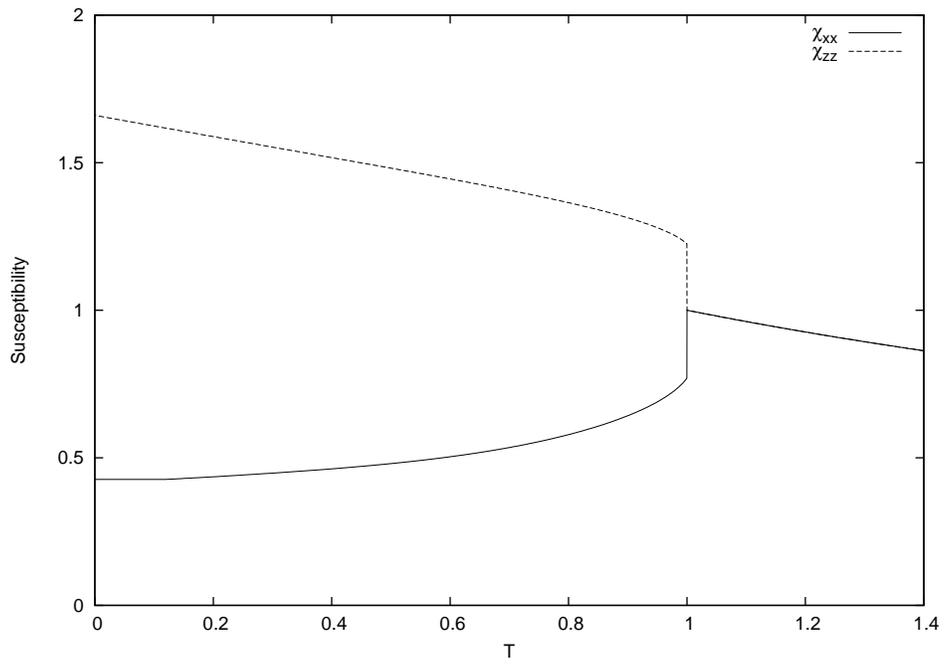}
\end{center}
\caption[Temperature dependence of the susceptibilities 
for $v_2 = -v_1$.]{
\label{fig5}
\interlinia
Temperature dependence of the susceptibilities
for $v_2=-v_1$ [half-line (f) in Fig. 1].
This value of the $v_2/v_1$ cannot be obtained from
the excluded volume method.
$T$ denotes the dimensionless temperature.}
\end{figure}

\begin{figure}
\begin{center}
\includegraphics{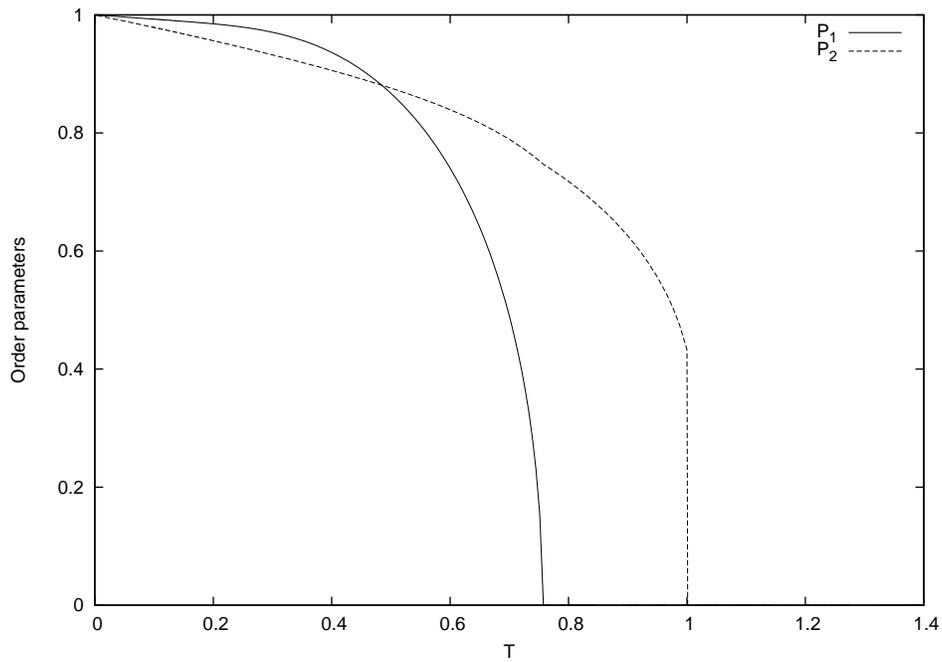}
\end{center}
\caption[Temperature dependence of the order parameters 
for $v_2=5v_1$.]{
\label{fig6}
\interlinia
Temperature dependence of the order parameter
$\langle P_1 \rangle$ and $\langle P_2 \rangle$ 
for $v_2=5v_1$ [half-line (c) in Fig. 1].
There is the first order transition from the isotropic to
the uniaxial nematic phase at $T=1$ and next, the second order
transition to the ferroelectric phase at $T=0.763$.
$T$ denotes the dimensionless temperature.}
\end{figure}

\begin{figure}
\begin{center}
\includegraphics{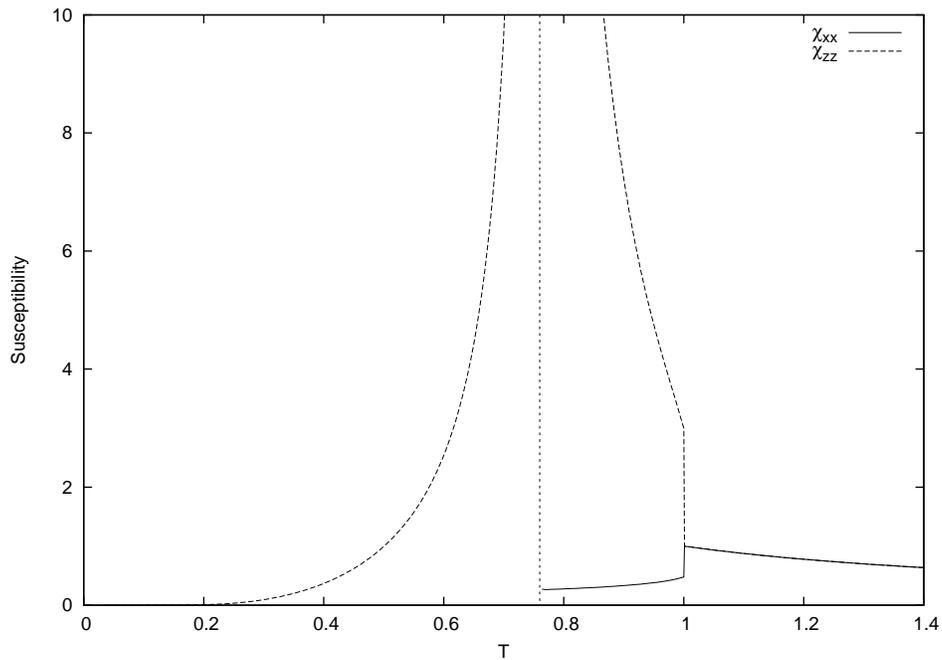}
\end{center}
\caption[Temperature dependence of the susceptibilities
for $v_2=5v_1$.]{
\label{fig7}
\interlinia
Temperature dependence of the susceptibilities
for $v_2=5v_1$ [half-line (c) in Fig. 1].
In the nematic phase we have $\chi_{xx}<\chi_{zz}$.
At the transition from the nematic to the ferroelectric 
phase $\chi_{zz}$ diverges (the Curie-Weiss law).
The vertical dashed line denotes the transition point between
nematic and ferroelectric phases. 
$T$ denotes the dimensionless temperature.}
\end{figure}

\begin{figure}
\begin{center}
\includegraphics{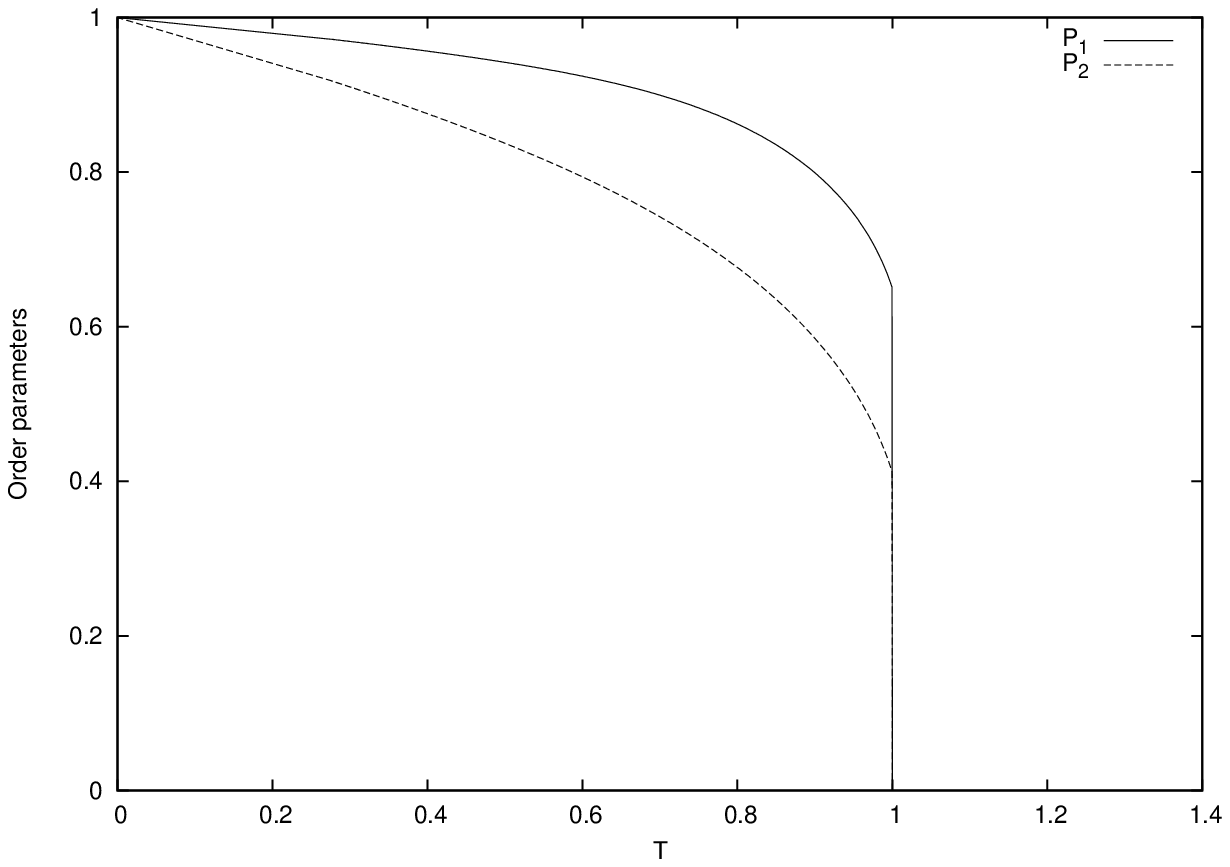}
\end{center}
\caption[Temperature dependence of the order parameters
for $v_2=v_1$.]{
\label{fig8}
\interlinia
Temperature dependence of the order parameters
$\langle P_1 \rangle$ and $\langle P_2 \rangle$ 
for $v_2=v_1$ [half-line (b) in Fig. 1].
There is the first order transition from the isotropic to
the ferroelectric phase.
$T$ denotes the dimensionless temperature.}
\end{figure}

\begin{figure}
\begin{center}
\includegraphics{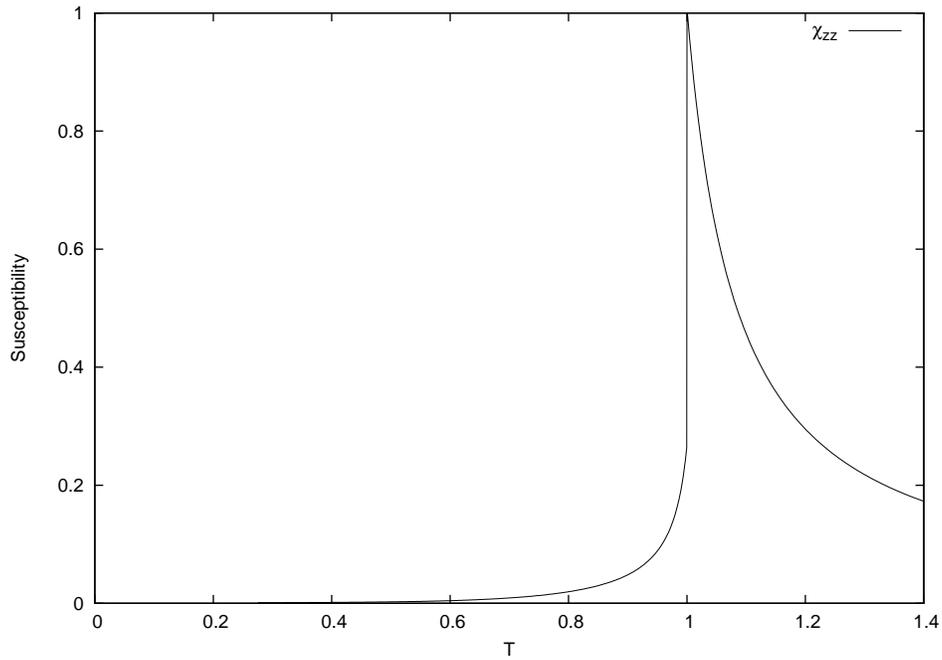}
\end{center}
\caption[Temperature dependence of the susceptibilities
for $v_2=v_1$.]{
\label{fig9}
\interlinia
Temperature dependence of the susceptibilities
for $v_2=v_1$ [half-line (b) in Fig. 1].
$\chi_{zz}$ is finite and it has the strong maximum at the 
transition from the isotropic to the ferroelectric phase.
$T$ denotes the dimensionless temperature.}
\end{figure}

\begin{figure}
\begin{center}
\includegraphics{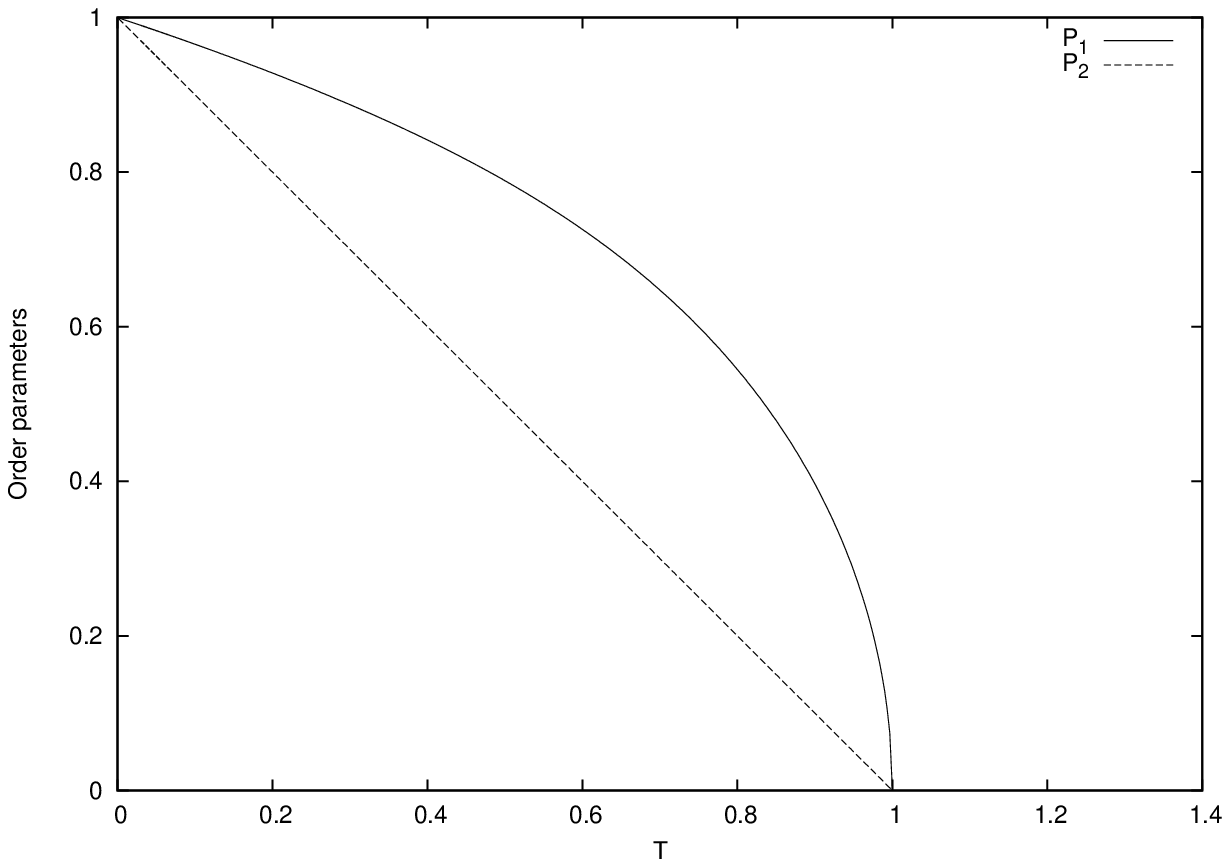}
\end{center}
\caption[Temperature dependence of the order parameters
for $v_2=0$.]{
\label{fig10}
\interlinia
Temperature dependence of the order parameters
$\langle P_1 \rangle$ and $\langle P_2 \rangle$ 
for $v_2=0$ [half-line (a) in Fig. 1].
There is the second order transition from the isotropic 
to the ferroelectric phase.
$T$ denotes the dimensionless temperature.}
\end{figure}

\begin{figure}
\begin{center}
\includegraphics{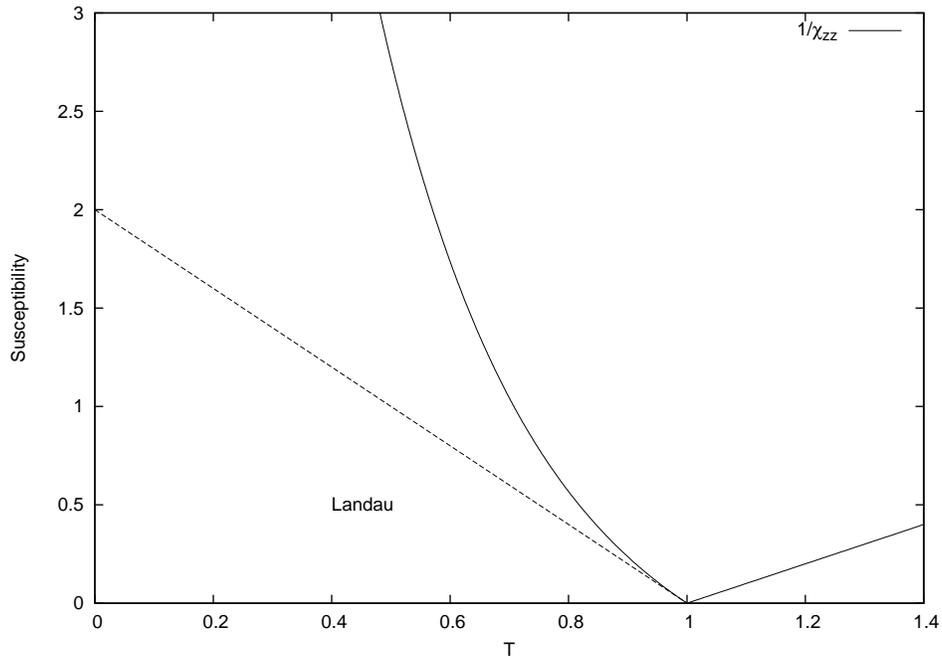}
\end{center}
\caption[Temperature dependence of the inverse of the 
susceptibilities for $v_2=0$.]{
\label{fig11}
\interlinia
Temperature dependence of the inverse of the susceptibilities
for $v_2=0$ [half-line (a) in Fig. 1].
The susceptibility $\chi_{zz}$ in the ferroelectric phase 
is half of that in the isotropic phase 
(in the neighbourhood of the transition point).
The dashed line shows the results from the Landau description.
$T$ denotes the dimensionless temperature.}
\end{figure}

\end{document}